\documentclass[aps,prb,twocolumn,showpacs,amsmath,amssymb]{revtex4}

\usepackage{graphicx}
\usepackage{dcolumn}
\usepackage{bm}
\usepackage[latin1]{inputenc}

\begin{document}

\title{Subharmonic gap structures and Josephson effect in MgB$_{2}/$Nb micro-constrictions}

\author{F. Giubileo}
  \email{giubileo@sa.infn.it}
\author{M. Aprili$^\dagger$}
\author{F. Bobba}
\author{S. Piano}
\author{A. Scarfato}
\author{A. M. Cucolo}
\affiliation{Physics Department and CNR-SUPERMAT Laboratory,
University of Salerno, Via S. Allende, 84081 Baronissi (SA), Italy
 \\ $^\dagger$ Laboratoire de Spectroscopie en
Lumi\`{e}re Polaris\'{e}e, ESPCI, \\ 10 rue Vauquelin, 75005
Paris, France \\ and CSNSM-CNRS, Bat. 108 Universit\'{e} Paris-Sud, 91405 Orsay, France
}%

\begin{abstract}
Superconducting micro-constrictions between Nb tips and high quality
MgB$_{2}$ pellets have been realized by means of a point-contact
inset, driven by a micrometric screw. Measurements of the
current-voltage characteristics and of the dynamical conductance
versus bias have been performed in the temperature range between 4.2
K and 500 K.  Above the Nb critical temperature T$_{C}^{Nb}$, the
conductance of the MgB$_2$/normal-metal constrictions behaves as
predicted by the BTK model for low resistance contacts while high
resistance junctions show quasiparticle tunneling characteristics.
Consistently, from the whole set of data we infer the value
$\Delta_{\pi} = 2.5 \pm 0.2$ meV for the three-dimensional gap of
MgB$_2$. Below T$_{C}^{Nb}$, low resistance contacts show Josephson
current and subharmonic gap structures (SGS), due to multiple
Andreev reflections. Simultaneous observations of both features,
unambiguously indicate coupling of the 3D band of MgB$_2$ with the
Nb superconducting order parameter. We found that the temperature
dependence of the Josephson critical current follows the classical
Ambegaokar-Baratoff behavior with a value $I_CR_N=(2.1 \pm 0.1)$ meV
at low temperatures.
\end{abstract}

\maketitle

\section{Introduction}\label{intro}
The discovery of superconductivity in the binary intermetallic
MgB$_{2}$ compound \cite{Akimitsu} has given rise to a considerable
effort in the condensed matter community in the last years. Besides
the great interest in understanding the new physics originating from
the multiband nature of this material, attention has been paid to
the attractive potential applications in superconducting
electronics, because of the remarkably high critical temperature
(T$_C\simeq 39\ K$) simple crystal structure, relatively long
coherence lengths \cite{Xu} ($\xi^c \simeq 25$\AA, $\xi^{ab} \simeq
65$\AA)
 and low surface resistance \cite{Lee} ($R_S \simeq 0.8 m\Omega$ at
T=24\, K).

From the structural point of view, the intermetallic MgB$_{2}$
superconductor is very similar to graphite with crystal lattice
formed by honeycomb layers of B atoms intercalated by layers of Mg
atoms, sitting at the center of each underlying hexagon. From the
theoretical point of view, this compound presents a rare example of
two disconnected parts of the Fermi surface: a two-dimensional (2D)
hole-type $\sigma$ bands, and a three-dimensional (3D) electron-type
$\pi$ bands \cite{Kortus,Shulga}. The unusual consequence of this
band structure makes that in the clean limit \cite{Liu} two
different energy gaps are formed at the Fermi level, both closing at
the same temperature T$_{C}$. Indeed, two superconducting energy
gaps have been experimentally observed by different techniques,
including tunneling spectroscopy \cite{Giubileo1,Iavarone,Suderow,
Szabo, Bougoslavski, Gonnelli, Schmidt,Carapella,Badr} specific heat
measurements \cite{Wang,Bouquet}, Raman spectroscopy
\cite{Chen,Quilty}, and high-resolution photoemission
\cite{Takahashi}. The majority of these studies agree with the
presence of a larger gap around $\Delta_{\sigma}$=7.0$\div$7.5 meV
attributed to the 2D $\sigma$-band and a smaller gap
$\Delta_{\pi}$=2.0$\div$2.8 meV due to the 3D $\pi$-band. In the
dirty limit a large amount of impurity scattering causes the two
gaps to merge to an intermediate value $\Delta_D \simeq$ 4.1 meV,
that closes at a reduced T$_{C}$ \cite{Brinkman,Fab}.

The physics of multiband superconductors has been intensively
studied since the appearance of the original theoretical works
\cite{Shun,Liu,Brinkman} and different phenomena are expected due to
the presence of different condensates in the same material. The
predictions have been tested in few cases, as e.g. in Niobium doped
SrTiO$_3$ \cite{Binning} or in Nickel borocarbides \cite{Nichel}.
Due to these considerations, MgB$_2$ appeared, from the beginning,
as a natural candidate to investigate peculiarities of a two band
superconductor. Recently, a multiband model for quasiparticle and
Josephson tunneling in MgB$_2$ based junctions has been developed
\cite{Brinkman}. Depending on the different bands exposed at the
sides of the insulating barrier, different temperature behaviors of
the Josephson current have been predicted with values of the
I$_C$R$_N$ product at low temperatures as high as 9.9 mV and 5.9 mV
for tunneling along the a-b planes and c direction, respectively. In
the experiments, however, effective Josephson coupling of the 2D
band $\sigma$ with a 3D band, both of a MgB$_2$ or of a conventional
superconducting counterelectrode, has been not observed. In addition
to this, the measured I$_C$R$_N$ values are often severely depressed
and regardless to the junction orientation, different temperature
dependencies of the Josephson current are often reported
\cite{Gonnelli, Carapella, Saito, Tao, Shimakage, Ueda}. Similar
behavior is expected in proximity coupled S--I--N--S structures
\cite{Barone}, and in MgB$_2$ junctions it has been attributed to
degradation of T$_C$ at the interface and/or to the barrier nature
and quality.

Besides the analysis of the Josephson coupling, the study of SGS at
voltages less than $\Delta$ is itself interesting. These resonances,
due to the coupling of the gap functions at the sides of the
barrier, have been mainly studied in symmetric S--I--S junctions
\cite{Taylor, Yason,ref11, pragati, ref19, Blonder, OBTK, Arnold,
1995}. In this case, conductance enhancements appearing at voltages
$V_n=2\Delta / n$ (n=1,2,3...) have been observed
\cite{a,physrev,Peshkin, 1994, Urbina, 2000}. In junctions with
dissimilar superconductors (S--I--S') conductance structures appear
at energies $(\Delta_S+\Delta_{S'})/m$, (with m= 1,3,5...),
$\Delta_{S'}/n$ and $\Delta_{S}/n$ (with n=1,2,3,...), depending on
the energy ratio $\Delta_{S'}/ \Delta_{S}$. \cite{Zimmerman, Hurd}

In this paper we address the problem of the behavior of both SGS and
Josephson current in  a multiband superconductor. Since both
phenomena depend on the multiplicity of the order parameter in the
two electrodes, the simultaneous presence of SGS and Josephson
effect allows a consistent cross check to verify the effective
coupling of different condensates at the sides of the tunnel
barrier. To investigate this aspect more deeply, we have realized
point contact (PC) junctions between MgB$_2$ pellets and Nb tips.
The aim of our study was to check for the Josephson coupling of the
Nb order parameter with both the 2D and/or 3D MgB$_2$ bands, this
aspect reinforced by the appearance of the related subharmonic gap
resonances. In the following we report, to the best of our
knowledge, the first detailed study of the temperature dependence of
both subharmonic gap structures and Josephson current observed in
asymmetric Nb-MgB$_{2}$ micro-constrictions. Current-Voltage
characteristic (I-V) and conductance spectra (dI/dV vs V) were
measured by using a home built point-contact apparatus, in which a
Nb tip was pressed into high quality MgB$_{2}$ pellets, to favor the
possible interaction with both bands of this compound.

\section{Superconducting contacts}

A point contact junction between a superconductor and a normal metal
(SN contact) or between two superconductors (SS' contact), is a
convenient geometry to study different aspects of superconductivity.
Indeed, the possibility of varying the strength of the potential
barrier, as well as of changing the contact area between the
electrodes, allows the observation of a large number of phenomena by
realizing a continuous variation from a metallic contact N-c-S (c is
the constriction), to a tunnel junction N-I-S (I is the insulating
barrier). This transition has been theoretically   and
experimentally studied by Blonder, Tinkham, and Klapwijk
(BTK)\cite{BTK} within a generalized semiconductor model, using the
Bogoliubov equations to treat the transmission and reflection of
quasiparticles at the interface. For a conventional
BCS-superconductor, there are two parameters in the model that are
varied to reproduce the conductance curves: the superconducting
energy gap $\Delta$ and a dimensionless parameter Z taking into
account for the barrier strength.

\vskip 0.2in

\begin{figure}[h]
\centering \includegraphics[width=8.50cm]{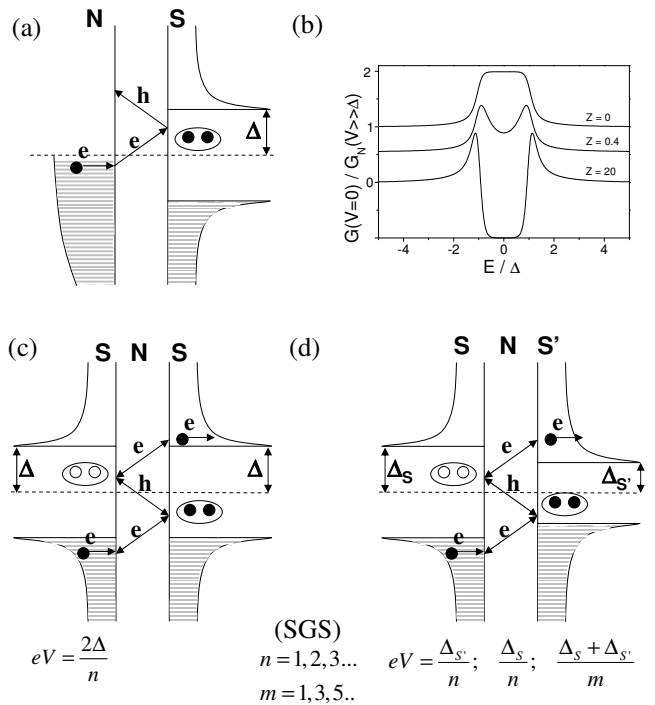}
\caption{\label{scheme} (a) An electron coming from the normal
electrode with energy smaller than the energy gap cannot enter the
superconductor. It is reflected as a hole, leaving an extra charge
2e in the superconducting condensate (Andreev reflection). At low
temperatures, the conductance spectra depend on the barrier strength
Z, as shown in (b). The scale refers to the curve for $Z=0$, the
other curves are shifted for clarity. In the case of superconducting
electrodes (c), an electron is reflected back or transmitted. For
voltages $eV \le \Delta$ the Andreev reflected hole can not find an
empty state in the left electrode but is Andreev reflected in turn
as an electron. For energies $2\Delta/3\leq eV <\Delta/2$ the
electron finds an empty state in the right electrode, three charges
are transferred in this process (Multiple Andreev Reflections). For
smaller voltages, MAR processes that transfer four or more electron
charges carry the current. (d) In asymmetric tunnel junctions the
conductance enhancements appear at energies $(\Delta_S +
\Delta_{S'})/m$, with $m$=1,3,5,..., $\Delta_S/n$ and
$\Delta_{S'}/n$, with $n$=1,2,3,...}
\end{figure}

SN metallic contacts with low barriers (Z $\rightarrow$ 0) show
enhanced current at bias voltages less than the superconducting
energy gap because Andreev reflections are the dominant transport
process at the interface \cite{BTK}. In this case, an incoming
electron from the normal metal with energy E $< \Delta$ cannot enter
in the superconducting electrode and is reflected as a hole in the
normal metal, simultaneously adding a Cooper pair to the condensate
in the superconductor (Fig. \ref{scheme} (a)). This process causes
an increase of the conductance around zero bias with a maximum ratio
G(V=0)/G$_N$(V$\gg \Delta$)=2. On the other hand, in the tunneling
regime (Z$\rightarrow \infty$), the probability of Andreev
reflections is negligible and the BTK model yields a conductance
characteristic that, at low temperatures for conventional
superconductors, directly reproduces the BCS superconducting density
of states at the Fermi level. In the intermediate case, both
tunneling and Andreev reflections contribute to the transport
through the barrier (Fig. \ref{scheme} (b)).

A different situation is realized when both electrodes forming the
micro-constriction are superconducting. The main features in the
current voltage characteristics of such junctions are the presence
of Josephson current and SGS. In particular, these last features
have been observed in junctions between identical superconductors as
conductance peaks appearing at bias voltages V=2$\Delta$/$ne$ with
$n$ = 1,2,3,... \cite{physrev,Peshkin, Urbina}. Different mechanisms
have been suggested to describe this phenomenon: initially,
correctly multiparticle tunneling model (MPT) \cite{Schrieffer} was
able to predict the voltage positions of the SGS. Successively, a
satisfactory explanation of the subgap structures has been proposed
in the BTK theory\cite{BTK} and then extended in the OBTK model by
Octavio et al. \cite{OBTK}: SGS are originating from multiple
Andreev reflections (MAR) at the superconductor--normal-metal
interfaces, see Fig. \ref{scheme}(c). This model has been then
generalized by Arnold \cite{Arnold} to any kind of junctions between
two superconductors and in particular to the superconducting point
contacts by using a modified tunneling Hamiltonian approach. It is
now understood that MPT model represents the low transparency
perturbation theory limit of the more general process of MAR
\cite{1995}.

Unlike symmetric junctions, very few reports  in the literature
address the behavior of asymmetric S--c--S' constrictions, see Fig.
\ref{scheme}(d), both  theoretically \cite{OBTK, Hurd} and
experimentally \cite{a, Zimmerman}. In particular, calculation of
the current-voltage relations for ballistic S--N--S' junctions by
means of transmission formalism \cite{Hurd} showed that, for finite
temperatures, SGS appear in the differential conductance at energies
$eV_n=(\Delta_S + \Delta_{S'})/m$,with $m$=1,3,5,...,
$eV_n=\Delta_S/n$, and $eV_n=\Delta_{S'}/n$, with $n$=1,2,3,...

\section{Conductance characteristics for $\mathbf{T >T_C^{Nb}}$}

The polycrystalline MgB$_{2}$ pellets used in the present work,
showed resistive superconducting transitions at T$_{C} (\rho = 0)$ =
38.8 K, with $\Delta $T$_{C} \simeq $ 0.5 K. The sample surface was
chemically etched in a 1\% HCl solution in pure ethanol. The Nb tips
were prepared by cutting a thin (0.2 mm) Nb wire then treated by
electrochemical etching. Soon after, the PC inset was placed in
liquid $^{4}$He to limit surface degradation effects. The contacts
were established by driving the Nb tip into the sample surface at
low temperatures. The vertical movement of the tip, driven by a
micrometric screw with a precision of about 0.1 $\mu$m, allowed the
tuning of the contact resistance from tunneling regime to metallic
contact. Our experimental setup resulted to be extremely stable,
showing no relevant effects of thermal contraction, so that in many
cases it was possible to vary the junction temperature without
affecting the contact geometry.

All measurements were performed in the temperature range between 4.2
K and 50 K. Current and dI/dV versus V characteristics were measured
by using a standard four-probe method and a lock-in technique by
superimposing a small ac-modulation to a slowly varying bias
voltage. Each measurement comprises two successive cycles in order
to check for the absence of heating hysteresis effects
\cite{hysteretic, hysteretic1}.

\begin{figure}[t!]
\centering \includegraphics[width=7.6cm]{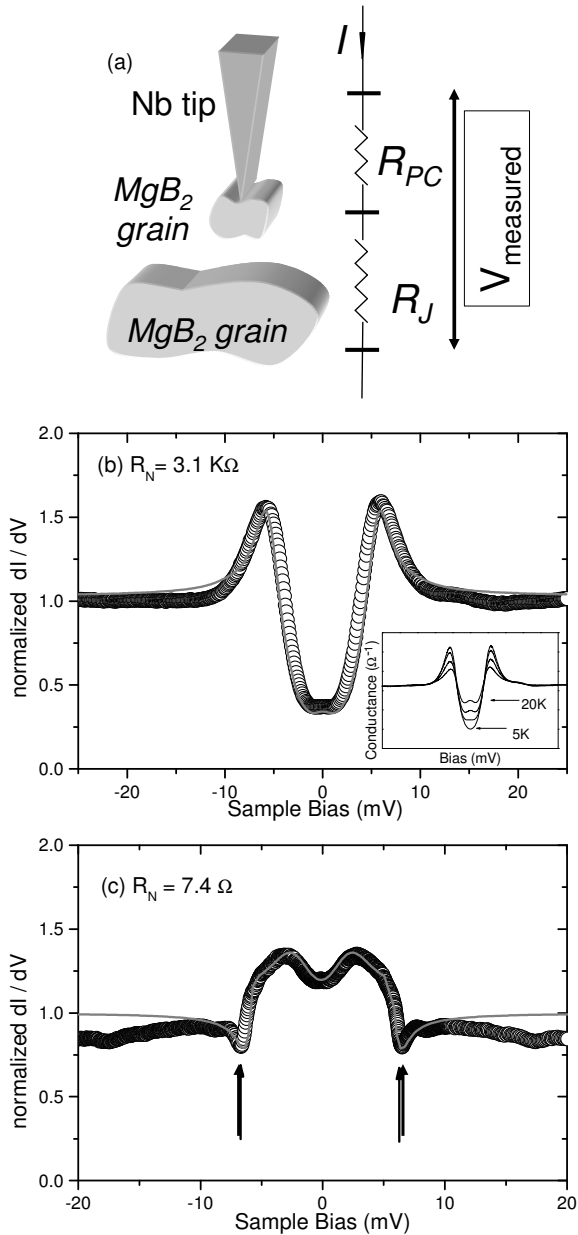}
\caption{\label{GVlowT}(a) A small MgB$_2$ grain remains on the Nb
tip realizing two junctions in series: the Point Contact between the
Nb tip and the small grain with resistance R$_{PC}$, and the
Josephson junction between the two MgB$_2$ grains with resistance
R$_J$. The measured voltage, V$_{meas}$, is the sum of the voltage
drops at both junctions. (b) Tunneling conductance for a high
resistance contact measured at T = 12 K. Experimental data
(scattered graph) are compared to the theoretical simulation (solid
line) for a S--I--S tunnel junction between two MgB$_2$ with
$\Delta_\pi$= 2.7 meV and $\Gamma$ = 0.7 meV . (c) Conductance
spectrum at T =12 K for a low resistance contact. The theoretical
simulation (solid line) has been obtained with $\Delta_\pi$= 2.4
meV, Z=0.66 and $\Gamma$ = 0  considering the formation of an
intergrain MgB$_2$/MgB$_2$ junction in series with $R_J=1.5 \Omega$
and $I_J=1.2mA$.}
\end{figure}

In Fig. \ref{GVlowT}(b), (c) we show the conductance spectra
measured for high and low junction resistances above the niobium
critical temperature, T$_C^{Nb}$. These are representative of
several measurements carried out on different contacts. High contact
resistances were obtained by pushing the Nb tip into the MgB$_2$ and
then slightly releasing the pressure. In this case, the conductance
curves show the typical tunneling behavior, Fig. \ref{GVlowT}(b).
The temperature dependence of the zero bias tunneling conductance,
reported in the inset, suggests that rather than a SIN junction, the
contact is formed between two MgB$_2$ grains. Indeed, a zero bias
conductance peak is found as for SIS junctions, due to thermally
activated quasiparticles with spectral weight increasing for rising
temperatures. The SIS behavior has been observed in other
experiments of point contact spectroscopy in polycrystalline HTC
superconductors \cite{Suderow, ZADA} and it has been attributed to a
small piece of the base electrode captured by the tip apex. Due to
the granularity of the samples, also in our case when releasing the
pressure, one MgB$_2$ grain remains on the Nb tip, see Fig.
\ref{GVlowT}(a). Two junctions in series are so formed and in the
analysis of the data attention has to be payed to the relative
weight of the related junction resistances. In the tunneling regime,
the contribution of the point contact junction is not critical,
since $R_{PC}<<R_J$. On the other hand, this has to be taken into
account, in the point contact regime, with both resistances of the
same order of magnitude. The tunneling conductance characteristic,
measured at T=12 K (scattered graph in Fig. \ref{GVlowT}(b)), were
so compared to the theoretical fitting for a S-I-S tunnel junction
between two identical superconductors (solid line). Since R$_N=3.1\
k\Omega$, we did not consider the contribution of the point contact
in series. $\Delta_\pi = 2.7 \, meV$ was used in the fitting in
which a broadening parameter, representative of the quasiparticle
finite lifetime, $\Gamma$ = 0.7 meV was also included \cite{Dynes}.

The conductance spectrum of a low resistance contact is shown in
Fig. \ref{GVlowT}(c), where the expected Andreev Reflection behavior
with conductance enhancement for eV$<\Delta$ is observed. The
conductance characteristic also shows dip structures at energies
higher than the AR feature (arrows in the figure), that are not
reproduced by the usual BTK theory. The origin of these dips has
been related to proximity effect \cite{strikers} and/or to the
formation of a junction in series \cite{cinesi}. In our case, for a
satisfactory fitting of the experimental data, it was necessary to
consider a more complex configuration, consistent with the
assumption shown in Fig. \ref{GVlowT}(a). Indeed, when the
resistance of the point contact junction (R$_{PC}$) is comparable
with the resistance of the Josephson junction (R$_J$), we have to
consider the formation of two junctions in series: the SN point
contact between the Nb-tip and MgB$_2$ and the inter-grain, low
resistance MgB$_2$/MgB$_2$ junction. As shown in Fig. \ref{GVlowT}
(a), as the PC tip/grain junction is approached to the base MgB$_2$
electrode, the latter junction reduces its resistance and the two
MgB$_2$ grains can be coupled by Josephson effect. The measured
voltage is now given by the sum of two contributions, i.e., the
voltage drop V$_{PC}$ at the point contact and V$_J$ at the
Josephson junction. In the limit of small capacitance C, the average
contribution V$_J$ can be simulated by the modified resistively
shunted junction model given by Lee\cite{Lee2}, where:
\begin{equation}
\langle V_J \rangle = \frac{2}{\gamma}R_J I_J
\frac{exp(\pi\gamma\alpha)-1}{exp(\pi\gamma\alpha)}T_{2}^{-1}
\end{equation}
with $\alpha=I/I_J$ the normalized current, I$_J$ the Josephson
current of the junction in  series, $\gamma=hI_J/eK_BT_n$ ($T_n$
being the effective noise temperature) and
\begin{equation*}
T_2= \int^{2\pi}_{0} d\varphi \sin\frac{\varphi}{2} I_1 \left(
\gamma \sin\frac{\varphi}{2} \right) exp \left[ -\left(
\frac{\gamma}{2} \alpha \right) \varphi \right]
\end{equation*}
where $I_1(x)$ is the modified Bessel function. The conductance
$\sigma (V)$ is then calculated by the formula:
\begin{equation}
 \sigma(V) = dI/dV = (dV_{PC}/dI + dV_J/dI)^{-1}\,,
\end{equation}
 where the PC
contribution is simulated by the usual BTK model. The best
theoretical fitting (solid line in Fig. \ref{GVlowT}(c)) reproducing
also the side deeps, was obtained for $\Delta_\pi = 2.4\ meV$ and $Z
= 0.66$, and no smearing factor $\Gamma$ was needed for this
junction.

We notice that the presence of the Josephson junction in series
introduces in the model two more fitting parameters, the resistance
$R_J=1.5 \Omega$  and the critical current I$_J$= 1.2 mA  of the
Josephson junction, these, however, are not independent one from the
other and from the measured resistance. We remark here that, since
the measured voltage V$_{meas}$=V$_{PC}$ + V$_J
>$ V$_{PC}$, a theoretical fitting that did not take into account
the presence of the additional V$_J$, would give an over-estimation
of the MgB$_2$ superconducting energy gap \cite{sitges, artsam}.
From our discussion, it appears that PC spectroscopy reveals two
type of junctions depending on the contact resistance:
MgB$_2$/MgB$_2$ at high resistances when the tip/grain is left far
from the surface and MgB$_2$/Nb at low resistances when the
tip/grain is pushed into the surface. From the whole set of data at
T$>$T$_{C}^{Nb}$, obtained in several locations, we have found as
average value of the 3D gap, $\Delta_{\pi}=2.5 \pm 0.2\, meV$.

\section{Conductance characteristics for $\mathbf{T <T_C^{Nb}}$}

\begin{figure}[h]
\centering \includegraphics{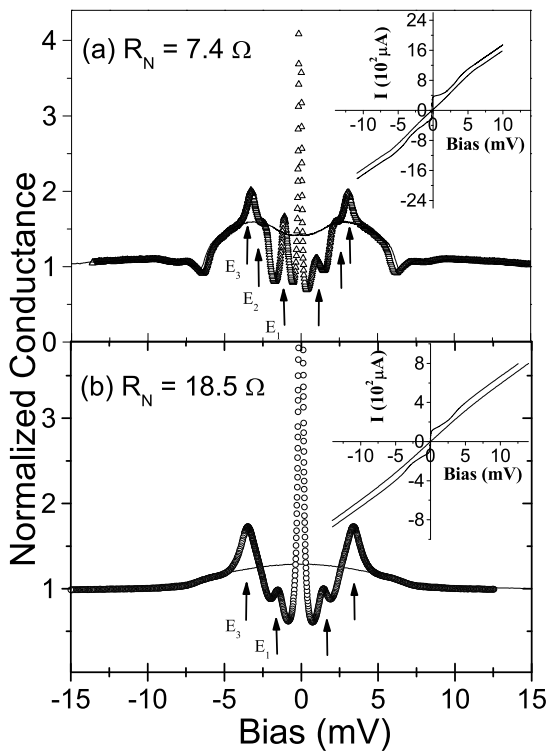}
\caption{\label{GVlowT2} Conductance spectra for two different
contacts measured at T = 7 K (scattered graphs). The conductance
measured at T = 12 K, i.e., above T$_C^{Nb}$, are also reported
(solid lines). The arrows indicate the energies of the SGS. Insets:
Current-Voltage characteristics of the same junctions.}
\end{figure}

As we already mentioned, the analysis of the Josephson coupling
between a two band superconductor and a conventional one has been
relatively less studied in the literature with contradictory results
about both the values of the I$_C$R$_N$ product and its temperature
dependence. In comparison with the theoretical predictions
\cite{Brinkman}, the majority of the reports indicate depressed
values of such product and different temperature behaviors
irrespectively to the weak link or tunnel nature of the junctions
\cite{Gonnelli, Carapella, Saito, Shimakage, Ueda}. To the best of
our knowledge, in the majority of the cases  coupling of the
conventional superconductor with only the 3D band of the multi-gap
material is inferred.

In this section we analyze the conductance data of low resistance
junctions for T$ <$ T$_C^{Nb}$ in which both Josephson effect and
subharmonic gap structures appear. Indeed, the simultaneous presence
of these features makes unambiguous any conclusion about the
coupling between the Nb order parameter and the 2D or 3D band of
MgB$_2$. In addition to this, the study of the behavior of the
subgap features in asymmetric junctions is itself of interest, since
few reports can be found in the literature.

In Fig. \ref{GVlowT2} (a), (b) we show the conductance spectra
measured at T=7 K for two different contacts.  The spectra are
characterized by the huge conductance peak at zero bias, signature
of the Josephson current flowing through the electrodes, as
confirmed by the corresponding I-V characteristics shown in the
insets. In addition to this, subharmonic gap structures at low
energies appear in both cases. Similar structures have been seldomly
observed in MgB$_2$ junctions, however a detailed analysis of their
origin and/or temperature dependence was not carried out
\cite{Suderow, ponomarev}.

\begin{figure*}[t!]
\centering \includegraphics[width=14cm]{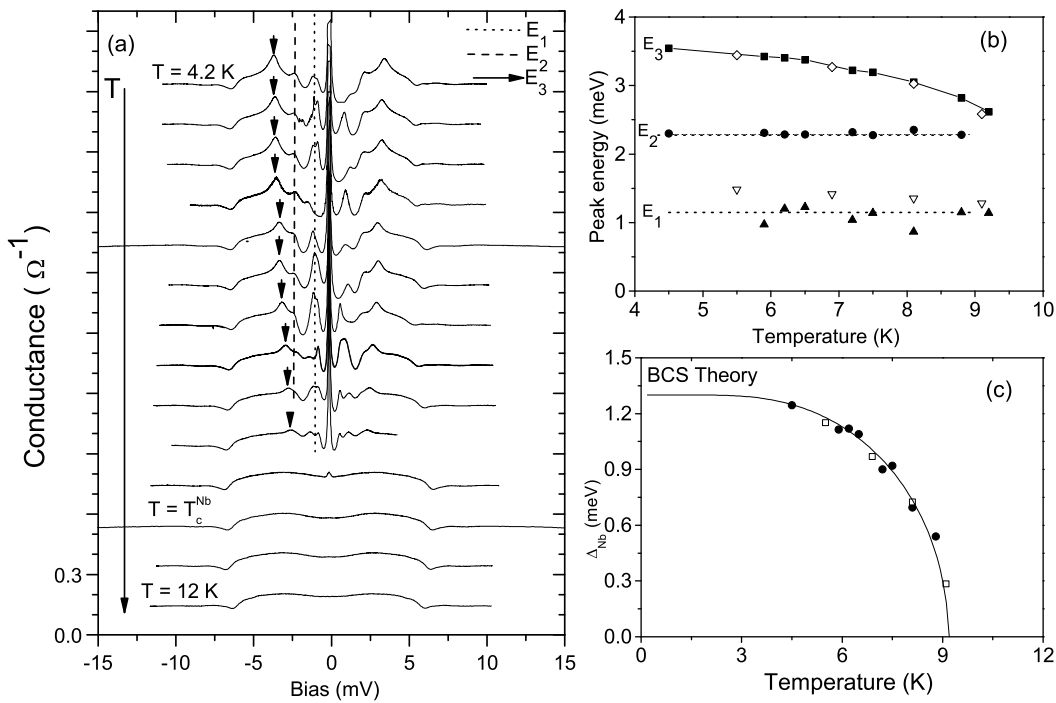}
\caption{\label{temp} (a) Temperature dependence of the conductance
spectra of the junction in Fig. \ref{GVlowT2} (a) in the range
between 4.2 K and 12 K. All the spectra have been shifted and the
scale is valid only for the highest temperature data. The arrows
indicate the energy positions of the E$_3$ feature. (b) Temperature
dependence of E$_1$, E$_2$, E$_3$. Dotted and solid lines are
guidelines for the eye. Open symbols correspond to sample in Fig.
\ref{GVlowT2} (b). (c) Temperature dependence of $\Delta^{Nb}$(T) =
E$_3$(T) - E$_2$(T), symbols, and theoretical curve, full line.}
\end{figure*}

In Fig. \ref{GVlowT2} we identify three principal features for
junction (a), localized at energies E$_1 \simeq \pm 1.2$ meV, E$_2
\simeq \pm 2.4$ meV, and E$_3 \simeq \pm 3.5$ meV, while only E$_1$
and E$_3$ are visible in junction (b), at the same energies. As
discussed above, these structures can be explained in terms of
multiple Andreev Reflections that enhance conductance at voltages
$\Delta_1/n$, $\Delta_2/n$, and ($\Delta_1+\Delta_2$). The most
useful way to identify the origin of various SGS can be obtained by
studying their temperature dependence. Due to the high stability of
our system, the dynamical conductance of both the contacts were
measured in the temperature range between 4.2 K and 12 K. Since the
two superconductors have quite different critical temperatures
(T$_{C}^{MgB_2}\simeq $39 K;\,  T$_{C}^{Nb}\simeq$ 9.2 K) for these
temperatures, only changes in the Nb energy gap are expected to
affect the evolution of the SGS.

In Fig. \ref{temp} (a) we show  the temperature dependence of the
conductance spectra for the junction of Fig. \ref{GVlowT2}(a). We
clearly see that the feature at the highest energy (E$_3$) changes
in temperature while E$_2 \simeq \pm 2.4$ mV and E$_1 \simeq \pm
1.2$ mV remain quite stable, suggesting that these are related to
MgB$_2$. In Fig. \ref{temp} (b), we report the temperature
dependence of the energy positions of the SGS for both junctions of
Fig. \ref{GVlowT2}, with solid and empty symbols referring to
junction (a) and (b), respectively. Due to the fact that all the
structures appear for voltages lower than 3.0 mV, we exclude the
influence of the 2D $\sigma$ band in the formation of the
resonances. We then make the following identification: E$_3
\rightarrow \Delta^{MgB_2}_{\pi} + \Delta^{Nb}$, E$_2 \rightarrow
\Delta^{MgB_2}_{\pi}$, and E$_1 \rightarrow \Delta^{MgB_2}_{\pi}/2$,
consistently with the MgB$_2$ value of $\Delta_\pi=2.4$ meV inferred
from the theoretical fitting of the data at T$ >$ T$_C^{Nb}$ (Fig.
\ref{GVlowT2} (c)). To confirm our hypothesis, we have extracted the
temperature dependence of the Nb gap from the data E$_3$(T) and
E$_2$(T), being $\Delta^{Nb}$(T) = E$_3$(T) - E$_2$(T). The result
is plotted in Fig. \ref{temp} (c) where the experimental data
correctly follow the theoretical behavior (full line) expected for
the Nb energy gap. In particular, the theoretical fitting gives
$\Delta^{Nb}(T=0) = 1.4\ meV$ and a local critical temperature
T$_C^{Nb}= $ 9.2 K. We notice that the E$_2$ structure only appears
in the lower resistance junction (Fig. \ref{GVlowT2}(a)), and its
amplitude is relatively depressed in comparison with E$_1$ and
E$_3$. Indeed, for this contact, we measured a higher value of the
Josephson current which, in comparison with junction (b), implies a
stronger coupling between the two electrodes.

\begin{figure}[h]
\centering \includegraphics[width=8cm]{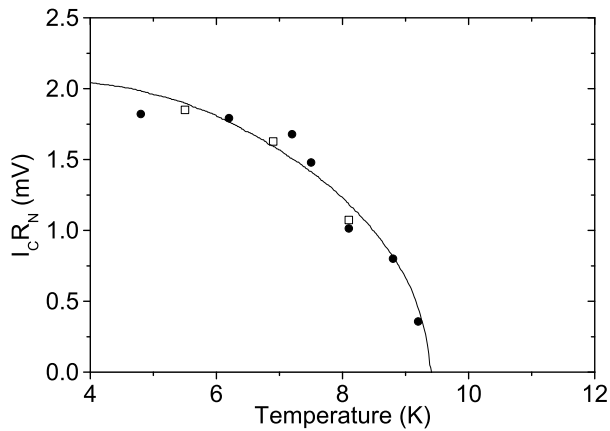}
\caption{\label{IcRn} Temperature dependence of the I$_C$R$_N$
product. The experimental data (scattered graph) are compared to the
Ambegaokar-Baratoff behavior (full line).}
\end{figure}

In Fig. \ref{IcRn}, we finally report the I$_C$R$_N$ product versus
temperature for both contacts in Fig. \ref{GVlowT2} (a) and (b).
I$_C=290\, \mu A$ and I$_C=110 \mu A$ were measured in these
junctions with R$_N=7.4 \Omega$ and R$_N=18.5 \Omega$, respectively.
This implies I$_C$R$_N= (2.1\pm 0.1)$ mV at T=4.5K, among the
highest values reported in the literature \cite{Carapella,Gonnelli,
Tao}. Our values are very close to those predicted when the
superconducting Nb couples with the MgB$_2$ $\pi$-band
\cite{Brinkman} and follow the expected Ambegaokar-Baratoff
temperature behavior \cite{Brinkman, Barone}. From our discussion it
unambiguously appears that coupling of the Nb energy gap only occurs
with the MgB$_2$ 3D $\Delta_{\pi}$ band, while in more than 50
measured contacts we never found evidence for coupling with the 2D
$\Delta_{\sigma}$ band. We have also demonstrated that it is in
principle possible to avoid the MgB$_2$ degraded surface layer,
responsible for both depressed values of the I$_C$R$_N$ product and
non conventional temperature behavior.

\section{Conclusion}

We have realized superconducting micro-constrictions between high
quality MgB$_2$ pellets and electrochemically etched Nb tips. The
conductance as a function of bias measured above the Nb critical
temperature reveals that an inter-grain MgB$_2$/MgB$_2$ junction is
often formed in series with the MgB$_2$/Nb contact. This results
from a small piece of MgB$_2$ remaining on the tip apex when
relieved from the pellet. Depending on the tip pressure the
MgB$_2$/MgB$_2$ contact resistance can be either larger (tip far
from the pellet) or comparable (tip into the pellet) with the
MgB$_2$/Nb point contact resistance. In the last case, an accurate
theoretical analysis has to be carried out to extract the correct
value of the MgB$_2$ superconducting energy gap. For T $<$
T$_C^{Nb}$, we have observed Josephson effect as well as subgap
resonances. We explain these features in terms of Subharmonic Gap
Structures due to Multiple Andreev Reflections. From the analysis of
the SGS, consistently with the values measured for T $>$ T$_C^{Nb}$,
we have extracted the correct temperature dependence of the Nb
energy gap and the value $\Delta_{\pi} \simeq 2.4$meV for the 3D
energy gap at the MgB$_2$ Fermi surface. In our junctions, at
T=4.5K, we have measured I$_C$R$_N$ values up to 2.2meV, among the
highest reported in literature. The temperature dependence of the
I$_C$R$_N$ product follows the classical Ambegaokar-Baratoff
behavior. Both observations completely confirm the results predicted
by a recent theoretical model \cite{Brinkman}. In addition to this,
the simultaneous observation of both Josephson current and SGS,
unambiguously indicate the coupling of the Nb energy gap with the
MgB$_2$ 3D band.

 \acknowledgments{\noindent M. Aprili acknowledges
Physics Department ``E.R. Caianiello'' of University of Salerno for
the kind hospitality.}


\begin{thebibliography}{99}

\bibitem{Akimitsu} J. Nagamatsu, N. Nakagawa, T. Muranaka, Y.
Zenitani, and J. Akimitsu, Nature \textbf{410}, 63 (2001).
\bibitem{Xu} M. Xu, H. Kitazawa, Y. Takano, J. Ye, k. Nishida, H. Abe, A.
Matsushita, N. Tsujii, and G. Kido, Appl. Phys. Lett. \textbf{79},
  2779 (2001).
\bibitem{Lee} S. Y. Lee, J. H. Lee, J. S. Ryu, J. Lim, S. H. Moon,
H. N. Lee, H. G. Kim, and B. Oh, Appl. Phys. Lett. \textbf{79},
3299 (2001).
\bibitem{Kortus} J. Kortus, I. I. Mazin, K. D. Belashchenko, V. P.
Antropov, and L. L. Boyer, Phys. Rev. Lett. \textbf{86}, 4656
(2001).
\bibitem{Shulga} S. V. Shulga, S.-L. Drechsler, H. Eshrig, H. Rosner, and W. E.
Pickett, cond-mat/0103154.
\bibitem{Liu}  A.Y. Liu, I. I. Mazin, and J. Kortus, Phys. Rev. Lett. \textbf{87},
087005 (2001).
\bibitem{Giubileo1} F. Giubileo,  D. Roditchev, W. Sacks, R. Lamy, D. X. Thanh, J.
Klein, S. Miraglia, D. Fruchart, J. Marcus, and P. Monod, Phys. Rev.
Lett. \textbf{87}, 177008 (2001).
\bibitem{Iavarone} M. Iavarone, G. Karapetrov, A. E. Koshelev, W. K. Kwok, G. W. Crabtree, D. G. Hinks,
W. N. Kang, E.-M. Choi, H. J. Kim, H.-J. Kim, and S. I. Lee, Phys.
Rev. Lett. \textbf{89}, 187002 (2002).
\bibitem{Suderow} P. Martinez-Samper, J. G. Rodrigo, G. Rubio-Bollinger, H. Suderow, S. Vieira, S. Lee, and S.
Tajima,
 Physica C \textbf{385}, 233 (2002).
\bibitem{Szabo} P. Szab\'{o}, P. Samuely, J. Kacmarc\'{i}k, T. Klein, J. Marcus, D. Fruchart, S. Miraglia, C. Marcenat, and A. G. M. Jansen,
Phys. Rev. Lett. \textbf{87}, 137005 (2001).
\bibitem{Bougoslavski} Y. Bugoslavsky, Y. Miyoshi, G. K. Perkins, A. V. Berenov, Z. Lockman,
 J. L. MacManus-Driscoll, L. F. Cohen, A. D. Caplin, H. Y. Zhai, M. P. Paranthaman, H. M. Christen, and M. Blamire,
 Supercond. Sci. Technol. \textbf{15}, 526 (2002).
\bibitem{Gonnelli} R. S. Gonnelli, A. Calzolari, D. Daghero, G. A.
Ummarino, V. A. Stepanov, G. Giunchi, S. Ceresara, and G. Ripamonti,
Phys. Rev. Lett. \textbf{87}, 097001 (2001).
\bibitem{Schmidt} H. Schmidt, J. F. Zasadzinski, K. E. Gray, and D. G. Hinks, Phys. Rev. Lett. \textbf{88}, 127002 (2002).
\bibitem{Carapella} G. Carapella, N. Martucciello, G. Costabile, C. Ferdeghini, V. Ferrando, and G.
Grassano,
 Appl. Phys. Lett. \textbf{80}, 2949 (2002).
\bibitem{Badr} M. H. Badr, M. Freamat, Y. Sushko, and K.-W. Ng, Phys. Rev. B \textbf{65}, 184516 (2002).
\bibitem{Wang} Y. Wang, T. Plackowski, and A. Junod, Physica C \textbf{355}, 179 (2001).
\bibitem{Bouquet} F. Bouquet, R.A. Fisher, N.E. Phillips, D.G. Hinks, and J.D.
Jorgensen, Phys. Rev. Lett. \textbf{87}, 047001 (2001).
\bibitem{Chen}  X.K. Chen, M.J. Konstantinovic, J.C. Irwin, D.D.
Lawrie, and J.P. Franck, Phys. Rev. Lett. \textbf{87}, 157002
(2001).
\bibitem{Quilty} J.W. Quilty, S. Lee, A. Yamamoto, and S.
Tajima,  Phys. Rev. Lett. \textbf{88}, 087001 (2002).
\bibitem{Takahashi}  T. Takahashi, T. Sato, S. Souma, T. Muranaka, and J. Akimitsu,
Phys. Rev. Lett. \textbf{86}, 4915 (2001).
\bibitem{Brinkman} A. Brinkman, A. A. Golubov, H. Rogalla, O. V. Dolgov, J. Kortus, Y. Kong, O. Jepsen, and O. K.
Andersen, Phys. Rev. B \textbf{65}, 180517(R) (2002).
\bibitem{Fab} F. Bobba, D. Roditchev, R. Lamy, E-M Choi, H-J Kim, W. N. Kang, V. Ferrando, C. Ferdeghini,
 F. Giubileo, W. Sacks, S-I Lee, J. Klein, and A. M. Cucolo,  Supercond. Sci. Technol. \textbf{16}, 167
 (2003).
 \bibitem{Shun} H. Suhl, B. T. Matthias, and L. R. Walker, Phys.
 Rev. Lett. \textbf{3}, 552 (1959).
\bibitem{Binning} G. Binnig, A. Baratoff, H. E. Hoenig, and J. G.
Bednorz, Phys. Rev. Lett. \textbf{45}, 1352 (1980).
\bibitem{Nichel} P. C. Canfield, P. L. Gammel, and D. J. Bishop, Phys. Today, \textbf{51},
40 (1998).
\bibitem{Saito} A. Saito, A. Kawakami, H. Shimakage, H. Terai, and Z. Wang
J. Appl. Phys. \textbf{92}, 7369 (2002).
\bibitem{Tao} H. -J. Tao, Z. -Z. Li, Y. Xuan, Z. -A. Ren, G. -C. Che, B. -R. Zhao and Z. -X.
Zhao, Physica C \textbf {386}, 569 (2003).

\bibitem{Shimakage} H. Shimakage, K. Tsujimoto, Z. Wang,
and M. Tonouchi, Appl. Phys. Lett. \textbf{86}, 072512 (2005).
\bibitem{Ueda} K. Ueda, S. Saito, K. Semba, T. Makimoto, and M.
Naito, Appl. Phys. Lett. \textbf{86}, 172502 (2005).


\bibitem{Barone} A. Barone and G. Patern\`{o}, \emph{Physics and Applications of the Josephson
effect} (John Wiley \& Sons, 1982).

\bibitem{Taylor} B. N. Taylor and E. Burstein, Phys. Rev. Lett.
\textbf{10}, 14 (1963).
\bibitem{Yason} I. K. Yanson, V. M. Svistunov, and I. M. Dmitrenko, Sov. Phys. JETP \textbf{21}, 650 (1965).
\bibitem{ref11} A. Griffin and J. Demers, Phys. Rev. B \textbf{4}, 2202
(1971).
\bibitem{pragati} P. Mukhopadhyay, J. Phys. F: Metal Phys.
\textbf{5}, 903 (1979).
\bibitem{ref19} C. J. Pethick and H. Smith,
Ann. Phys. (N.Y.) 119, 133 (1979).
\bibitem{Blonder} T. M. Klapwijk, G. E. Blonder, and M. Tinkham,
Physica B+C \textbf{109-110}, 1657 (1982).
\bibitem{OBTK} M. Octavio, M. Tinkham, G. E. Blonder, and T. M. Klapwijk, Phys. Rev. B \textbf{27}, 6739 (1983).
\bibitem{Arnold} G.B. Arnold, J. Low Temp. Phys. \textbf{68}, 1 (1987).
\bibitem{1995} E. N. Bratus, V. S. Shumeiko, and G. Wendin, Phys.
Rev. Lett. \textbf{74}, 2110 (1995).

\bibitem{a} J.M. Rowell and W.L. Feldman, Phys. Rev. \textbf{172},
393 (1968).
\bibitem{physrev} L. J. Barnes, Phys. Rev. \textbf{184}, 434 (1969).
\bibitem{Peshkin} M. A. Peshkin and R. A. Buhrman, Phys. Rev. B \textbf{28}, 161 (1983).
\bibitem{1994} N. van der Post, E. T. Peters. I. K. Yason, and J. M.
van Ruitenbeek, Phys. Rev. Lett. \textbf{73}, 2611 (1994).
\bibitem{Urbina} E. Scheer, P. Joyez, D. Esteve, C. Urbina, and M. H.
Devoret, Phys. Rev. Lett. \textbf{78}, 3535 (1997).
\bibitem{2000} B. Ludoph, N. van der Post, E. N. Bratus, E. V.
Bezuglyi, V. S. Shumeiko, G. Wendin and J. M. van Ruitenbeek, Phys.
Rev. B, \textbf{61}, 8561 (2000).

\bibitem{Zimmerman} U. Zimmenrmann, S. Abens, D. Dikin, K. Keck, and
V. M. Dmitriev, Z. Phys. Cond. Mat. \textbf{97}, 59 (1995).
\bibitem{Hurd} M. Hurd, S. Datta and P. F. Bagwell, Phys. Rev. B.\textbf{54}, 6557 (1996).

\bibitem{BTK} G. E. Blonder, M. Tinkham, and T. M. Klapwijk, Phys. Rev. B \textbf{25}, 4515 (1982).
\bibitem{Schrieffer} J. R. Schrieffer and J. W. Wilkins, Phys. Rev. Lett.
\textbf{10}, 17 (1963).
\bibitem{hysteretic} Yu. G. Naidyuk and I. K. Yanson, J. Phys. :
Condens. Matter \textbf{10}, 8905 (1998).
\bibitem{hysteretic1} K. Gloos, Phys. Rev. Lett. \textbf{85}, 5257
(2000).
\bibitem{ZADA} N. Miyakawa, P. Guptasarma, J. F. Zasadzinski, D. G. Hinks, and K. E. Gray, Phys. Rev. Lett. \textbf{80}, 157 (1998).
\bibitem{Dynes} R. C. Dynes, V. Narayanamurti, and J. P. Garno, Phys.
Rev. Lett. \textbf {41}, 1509 (1978).

\bibitem{strikers} G. J. Strijkers, Y. Ji, F. Y. Yang, C. L. Chien, and
J. M. Byers, Phys. Rev. B \textbf{63}, 104510 (2001).

\bibitem{cinesi} L. Shan, H. J. Tao, H. Gao, Z. Z. Li, Z. A. Ren, G. C. Che, and H. H. Wen, Phys. Rev. B \textbf{68}, 144510 (2003).

\bibitem{Lee2} P.A. Lee, J. Appl. Phys. \textbf{42}, 325 (1971).
\bibitem{sitges} S. Piano, F. Bobba, F. Giubileo, A. Vecchione, and A. M. Cucolo, accepted on J. Phys. Chem. Solids
(2004).
\bibitem{artsam} S. Piano, F. Bobba, F. Giubileo, A. Vecchione, M. Gombos and A. M. Cucolo, cond-mat/0509346.
\bibitem{ponomarev} Ya. G. Ponomarev, S. A. Kuzmichev, M. G. Mikheev, M. V. Sudakova,
S. N. Tchesnokov, N. Z. Timergaleev, A. V. Yarigin, E. G. Maksimov,
S. I. Krasnosvobodtsev, A. V. Varlashkin, M. A- Hein, G. M\"{u}ller,
H. Piel, L. G. Sevastyanova, O. V. Kravchenko, K. P. Burdina, and B.
M. Bulychev, Solid State Communications \textbf{129}, 85 (2004).


\end{thebibliography}
\end{document}